\documentclass[a4paper]{jpconf}
\usepackage{graphicx}
\usepackage{xcolor}
\usepackage{lineno}
\usepackage{hyperref}
\usepackage{xspace}
\usepackage{amssymb}
\usepackage{amsmath}

\begin{document}


\title{A look into the ``hedgehog'' events in pp collisions using a new event shape - {\it \textbf {flatenicity}}}

\author{Antonio Ortiz and Guy Paic}

\address{Instituto de Ciencias Nucleares, Universidad Nacional Aut\'onoma de M\'exico}

\ead{antonio.ortiz@nucleares.unam.mx, guypaic@nucleares.unam.mx}

\begin{abstract}
The UA1 and CDF collaborations have reported the presence of events with a very extended structure of low momentum tracks filling in a uniform way the $\eta-\varphi$ phase space. However, these events have not caused  interest with respect to the details of the interaction that originate them. We have undertaken to study whether such events are predicted by present event generators like \textsc{PYTHIA}~8. We report  the existence of events without any discernible jet structure in high-multiplicity pp collisions.  In the simulations those events originate from a moderate number of parton-parton scatterings within the same pp collision. A new event shape, $\rho$ ($flatenicity$), is proposed to allow for easily triggering on such events. 
\end{abstract}

\section{Introduction}

The proton-proton (pp) collisions at ultra-relativistic energies have shown very important similarities with heavy-ion collisions~\cite{Busza:2018rrf} like collectivity~\cite{CMS:2016fnw}, and strangeness enhancement~\cite{ALICE:2016fzo}. The origin of these effects in small systems like pp collisions is still a matter of debate~\cite{Nagle:2018nvi}. The high multiplicity pp collisions offer a very important laboratory to study high-energy density features of quantum chromodynamics (QCD). The present work illustrates the possibility to observe very rare pp collisions, and study their properties. These collisions would not, because of their rarity, give a noticeable contribution to the traditional measurements performed as a function of the mean charged particle multiplicity~\cite{ALICE:2018pal}.

It is interesting that before the LHC era in the long history of investigation of pp collisions there were, to our knowledge, only two mentions of ``strange'' pp collisions, one by the UA1 collaboration, and the other by the CDF experiment. The specificity and rarity  of these events lays in fact that the distribution of charged particles in the available phase space presents an almost isotropic distribution of low transverse momentum particles over a large pseudorapidity ($\eta$) range. Regarding the CDF document, C. Quigg  has named them as ``hedgehog'' events~\cite{Quigg:2010ew}. Some attempts to characterize these events involved the development of tools like event shapes~\cite{OrtizVelasquez:2009pey,Ortiz:2011pu}. Early LHC measurements of transverse sphericity as a function of the charged particle multiplicity unveiled features which suggested that a very active underlying event was needed by the event generators in order to explain  the properties of pp collisions with ${\rm d}N_{\rm ch}/{\rm d}\eta>30$~\cite{ALICE:2012cor,Ortiz:2017jho}. In other words, high multiplicity pp collisions seemed to be more isotropic (hedgehog-like structure) than predicted by MC generators. Recently, the average transverse momentum as a function of charged particle multiplicity in isotropic events was found to be smaller than that measured in jet-like events~\cite{ALICE:2019dfi}. The preliminary ALICE results on identified particle $p_{\rm T}$ spectra show a clear event shape dependence which hints to a larger strangeness enhancement in isotropic events relative to jet-like events~\cite{Nassirpour:2020owz}. Other results as a function of the event classifier $R_{\rm T}$~\cite{Martin:2016igp,Ortiz:2017jaz} point in the same direction. However, given the limited acceptance ($|\eta|<0.8$) used to measure both the event multiplicity and event shape the sample can be biased towards pp collisions with large activity from hard radiation which would go to the transverse region of the di-hadron correlations~\cite{Bencedi:2021tst}. This bias has to be taken into consideration for the interpretation of the results.  However, in view of the plans of a new detector for heavy-ion physics at CERN, ALICE 3~\cite{ALICE:2803563}, it is pertinent to explore the hedgehog-event tagging in the proposed detector which would have excellent tracking capabilities at low momenta ($p_{\rm T}>0.1$\,GeV/$c$) within a wide acceptance ($|\eta|<4$).     

\begin{figure}[ht!]
\centering
\includegraphics[scale=0.32]{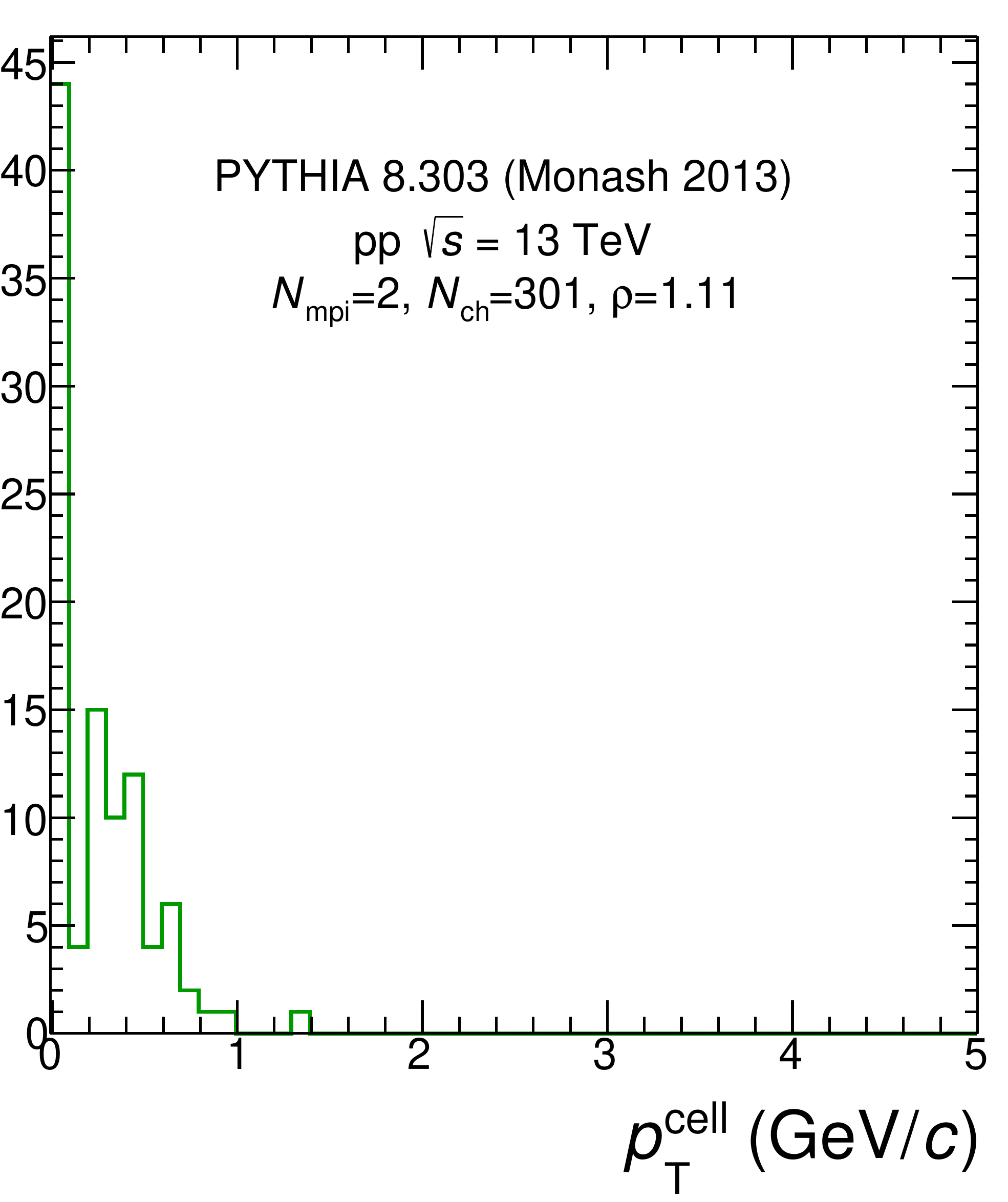}
\includegraphics[scale=0.32]{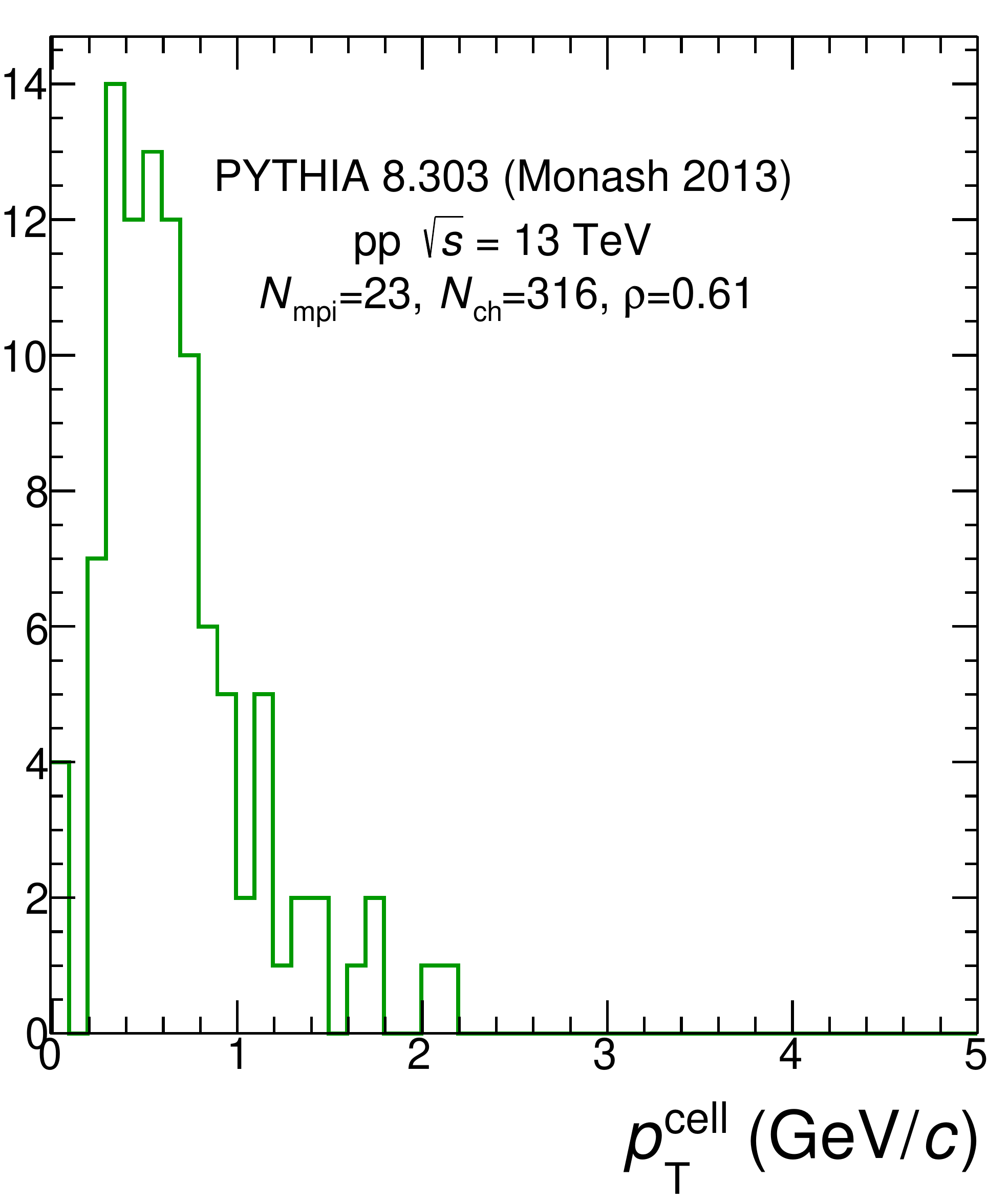}
\caption{Average transverse momentum in each cell for two high multiplicity pp collisions at $\sqrt{s}=13$\,TeV. Examples for events with low (left) and high (right) $N_{\rm mpi}$ are displayed. 
}
\label{fig:0}
\end{figure}

In the present paper we show that the \textsc{PYTHIA}~8\cite{Sjostrand:2014zea} Monte Carlo event generator predicts the existence of ``hedgehog'' events. In the model, this event topology originates from a moderate number of semi-hard parton-parton scatterings, termed as multiparton interactions (MPI), occurring within the same pp collision~\cite{PhysRevD.36.2019}. Together with the fragments from the beam-beam remnants, particles from MPI define the underlying event~\cite{ALICE:2019mmy}. The existence of such events in the event generator has opened the possibility to present a study of their properties and propose a potential way to tag them. Derived from these studies, we present a novel event structure parameter $\rho$-$flatenicity$ that permits identifying/triggering very easily the events and study their characteristics in future experiments like ALICE 3. But, the ideas can be adapted using the current detector capabilities of experiments at the LHC. 


\section{Flatenicity, $\rho$}

Hedgehog events are expected to have a uniform distribution of tracks with relatively low transverse momenta over the whole $\eta-\varphi$ range. In order to determine how uniform the particles' transverse momentum ($p_{\rm T}$) is distributed in a given event, the whole phase space is divided into 80 elementary cells. Given the tracking capabilities of the proposed ALICE 3, charged particles within $|\eta|<4$ and $p_{\rm T}>0.15$\,GeV/$c$ are considered in the calculation of $flatenicity$. In each cell,  the average transverse momentum is calculated ($p_{\rm T}^{\rm cell}$). Event-by-event, the distribution of $p_{\rm T} ^{\rm cell}$ (see e.g. Fig.~\ref{fig:0}) is used to get $flatenicity$ as follows:

\begin{equation}
\rho=\frac{\sigma_{p_{\rm T}^{\rm cell}}}{\langle p_{\rm T}^{\rm cell} \rangle},
\end{equation} 

which is the relative standard deviation of the $p_{\rm T}^{\rm cell}$ distribution. Events with jet signals on top of the underlying event are expected to have a large spread in the $p_{\rm T}^{\rm cell}$ values, the opposite is expected in the case in which particles with lower momenta would be isotropically distributed. Although, only results considering the average transverse momentum are presented, it is worth mentioning that average charged particle multiplicity can be used instead. Moreover, the method was also tested considering the acceptance covered with the detectors of ALICE 2~\cite{ALICE:2014sbx}.

\begin{figure}[ht!]
\centering
\includegraphics[scale=0.59]{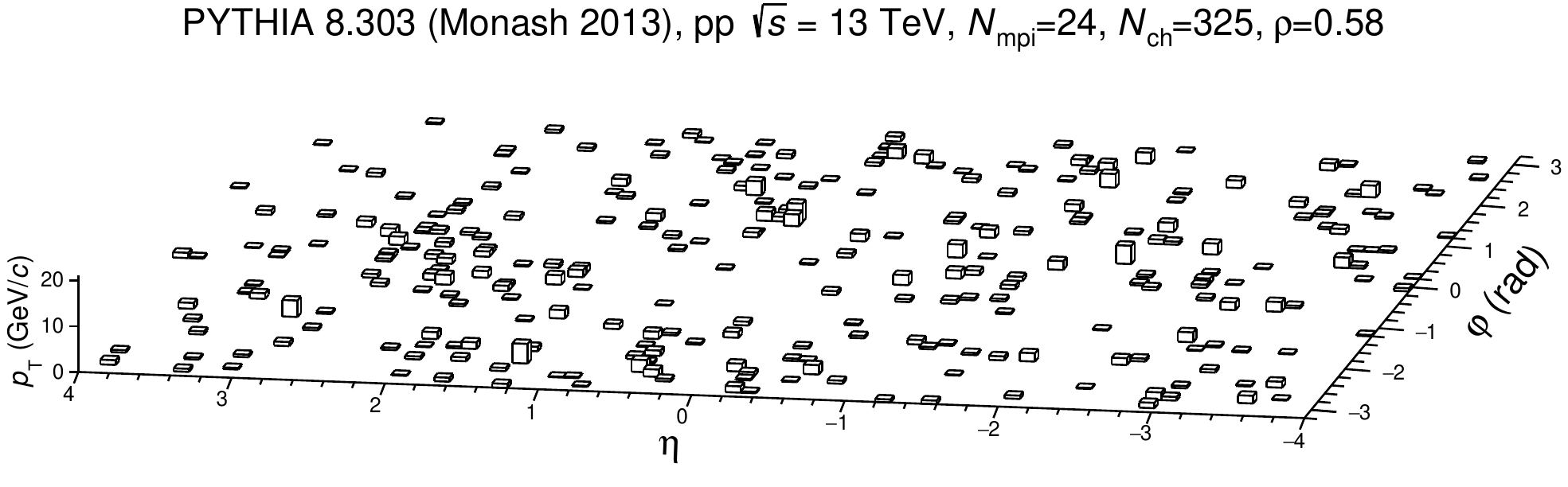}
\includegraphics[scale=0.59]{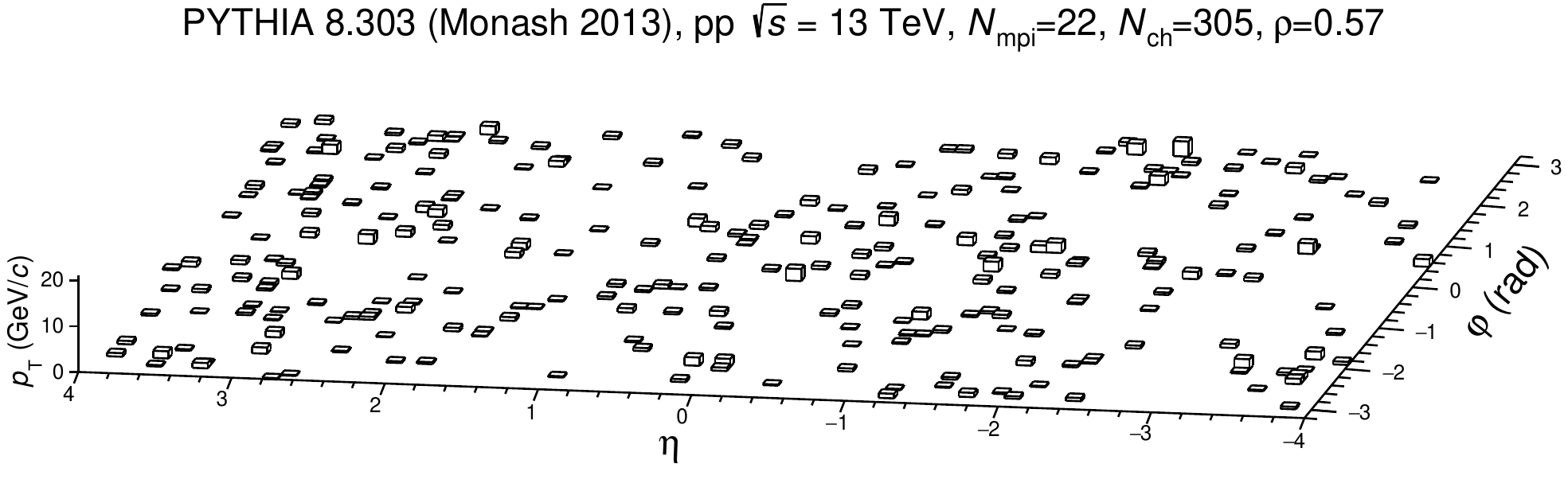}
\caption{Two event displays of hedgehog events generated with PYTHIA~8.244 tune Monash. More than 300 primary charged particles and more than 20 multiparton interactions were required within the same pp collision. 
}
\label{fig:1}
\end{figure}

In order to illustrate how $flatenicity$ works for hedgehog tagging, pp collisions at $\sqrt{s}=13$\,TeV were simulated using PYTHIA~8.244 (tune Monash). In this generator, hedgehog events originate from pp collisions where a moderate number of semi-hard scatterings occur within the same pp collision.  The $flatenicity$ values for this type of events is shown in Fig.~\ref{fig:1}, which exhibits the characteristic transverse momentum as a function of $\eta$ and $\varphi$ for hedgehog events.  Generally, the value $\rho=1$ looks like a limit between events with at least one well identified jet and ``hedgehog'' events, the latter occurring for values of $\rho <1$.

\begin{figure}[ht!]
\centering
\includegraphics[scale=0.59]{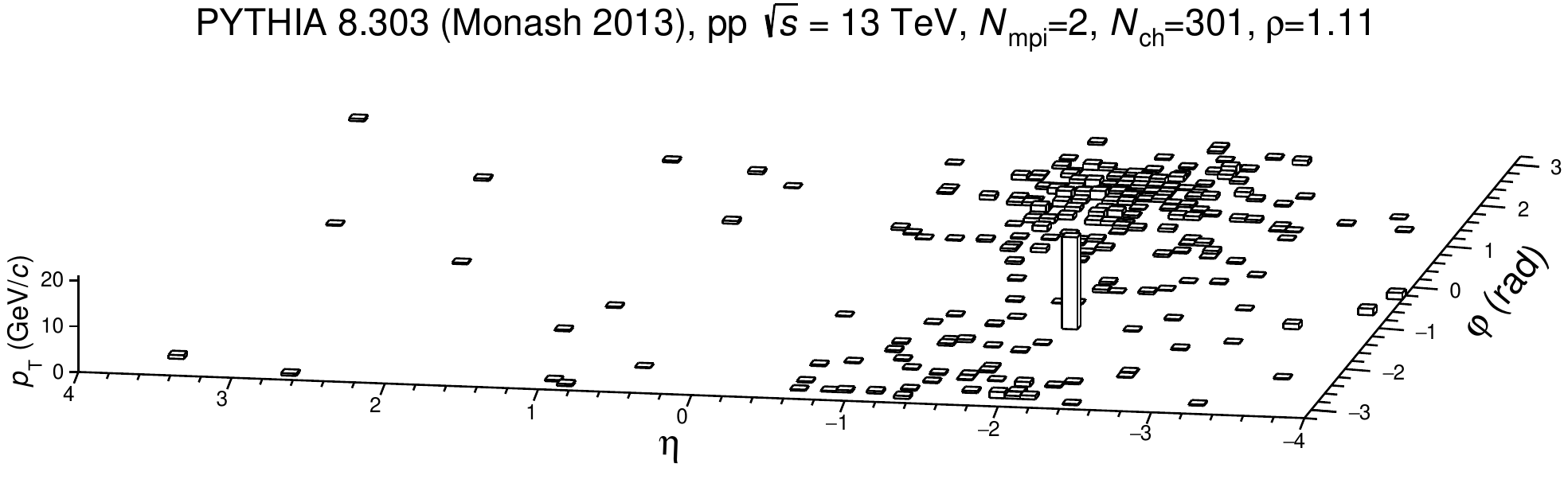}
\includegraphics[scale=0.59]{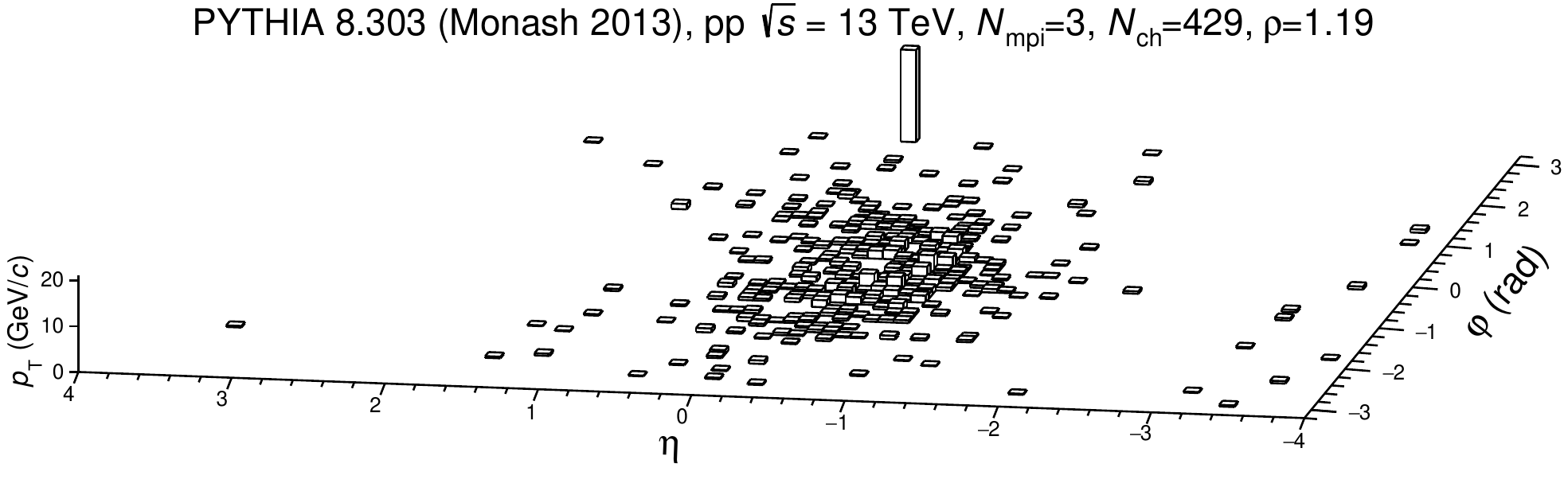}
\caption{Two event displays of events generated with PYTHIA~8.244 tune Monash. More than 300 primary charged particles and less than 5 multiparton interactions were required within the same pp collision. 
}
\label{fig:2}
\end{figure}

Events with similar multiplicity but originated from the fragmentation of less than five semi-hard scatterings more likely have $flatenicity$ larger than one,  their characteristic event structure is shown in Fig.~\ref{fig:2}.  In this case, we clearly see a jet signal which is accompanied by the recoil jet fragmenting into several charged particles, and also particles from a few additional multiparton interactions. 

\begin{figure}[ht!]
\centering
\includegraphics[scale=0.59]{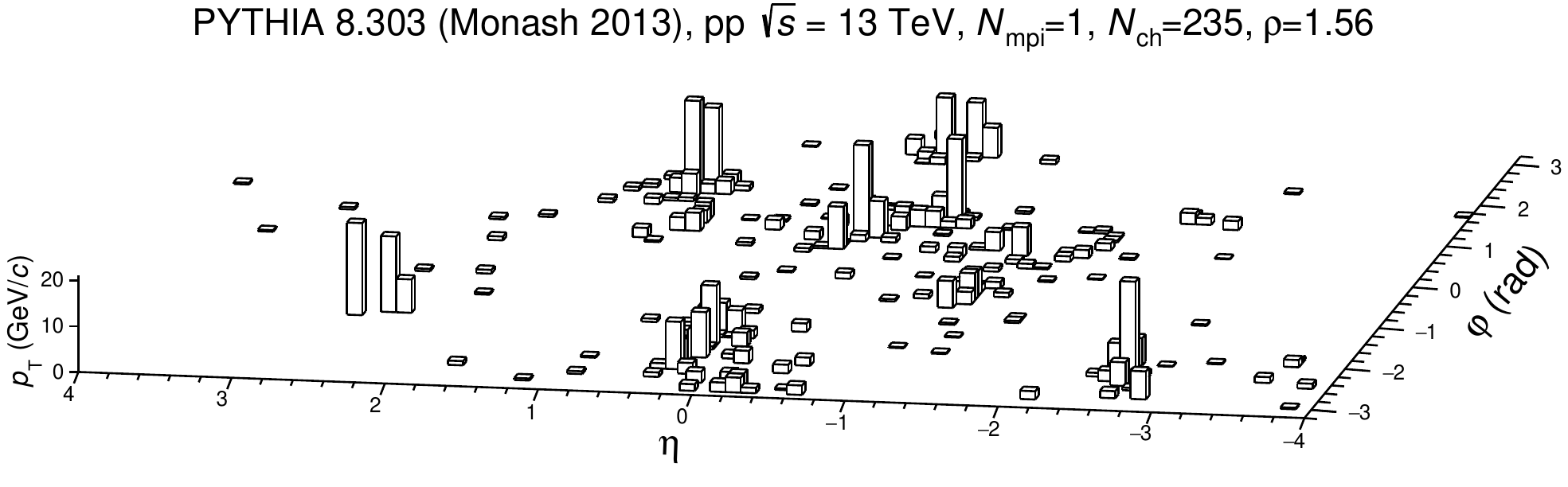}
\includegraphics[scale=0.59]{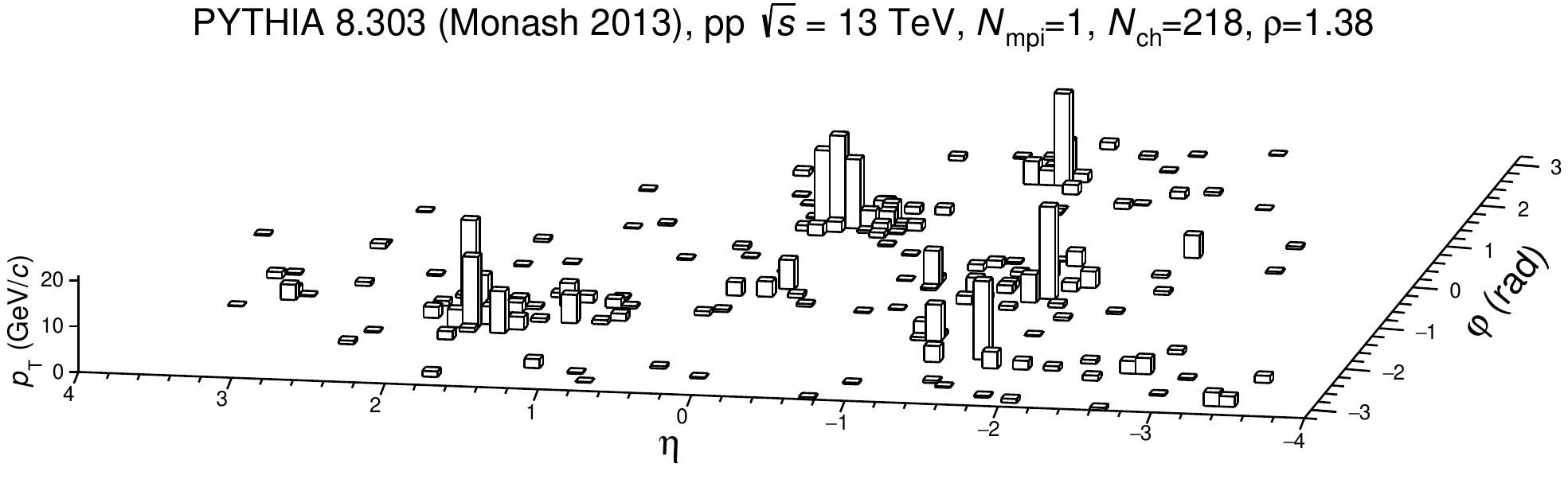}
\caption{Two event displays of events generated with PYTHIA~8.244 tune Monash. More than 200 primary charged particles from pp collision simulated without MPI are displayed. 
}
\label{fig:3}
\end{figure}

In order to illustrate the scenario without MPI, very high multiplicity pp collisions where generated in PYTHIA with the option PartonLevel:MPI=off. The event displays shown in Fig.~\ref{fig:3} clearly exhibit multi-jet structures, which originate $flatenicity$ values larger than 1. Even in this case, the event classifier $flatenicity$ seems to be robust to discriminate hedgehog events from multi-jet topologies.

\section{Flatenicity distributions}
\begin{figure}[ht!]
\centering
\includegraphics[scale=0.8]{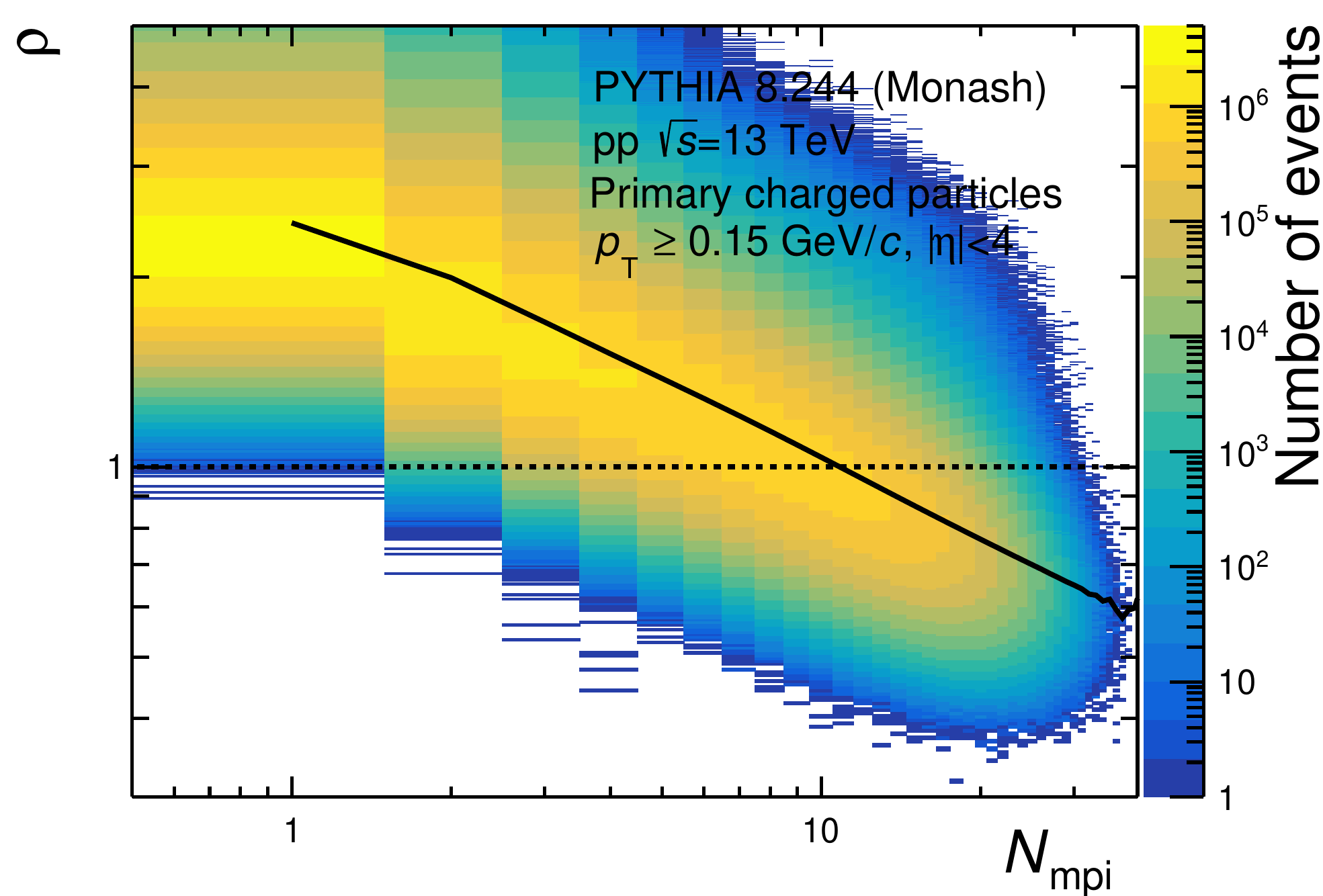}
\caption{Flatenicity as a function of the number of multiparton interactions, the number of events in each bin is depicted in different colors. The results were produced using PYTHIA~8.244 simulations of pp collisions at $\sqrt{s}=$13\,TeV. Only primary charged particles within $|\eta|<4$ and $p_{\rm T}\geq0.15$\,GeV/$c$ were considered.
}
\label{fig:4}
\end{figure}

\begin{figure}[ht!]
\centering
\includegraphics[scale=0.8]{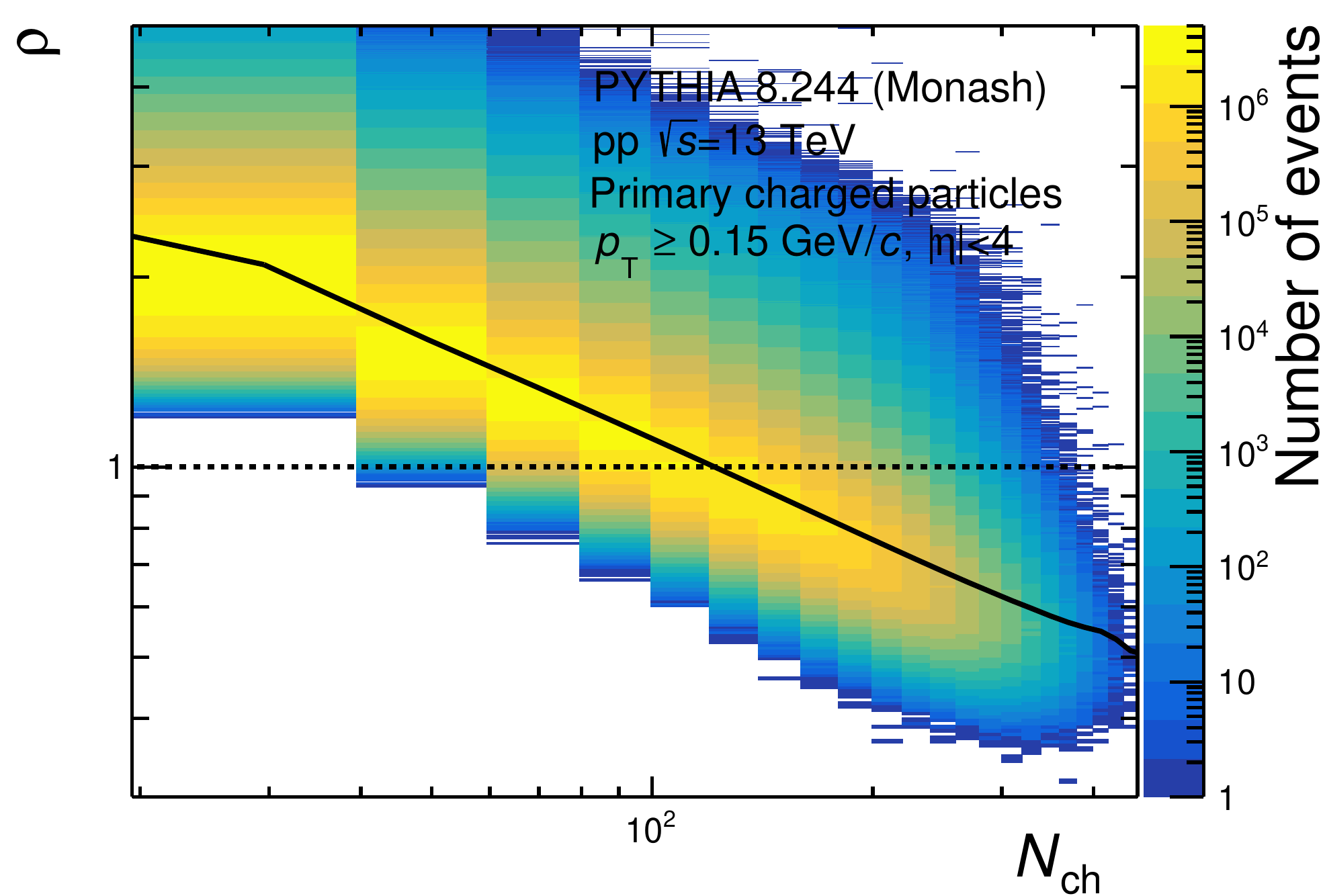}
\caption{Flatenicity as a function of the charged particle multiplicity, the number of events in each bin is depicted in different colors. The results were produced using PYTHIA~8.244 simulations of pp collisions at $\sqrt{s}=$13\,TeV. Only primary charged particles within $|\eta|<4$ and $p_{\rm T}\geq0.15$\,GeV/$c$ were considered.
}
\label{fig:5}
\end{figure}

The correlation between $flatenicity$ and the number of multiparton intactions is presented in Fig.~\ref{fig:4}. At low $N_{\rm mpi}$ the {\it flatenicity} distribution is very wide, it gets narrowed with increasing $N_{\rm mpi}$. The average {\it flatenicity} is significantly above unity for low $N_{\rm mpi}$, and goes below one for $N_{\rm mpi}>12$. For $N_{\rm mpi}>25$ the average {\it flatenicity} approaches to 0.5. By construction this behaviour is expected, the pp collisions with extremely large underlying event ($N_{\rm mpi}>25$) must have a uniform distribution of particles in the $\eta-\varphi$ space. Since in the experiment, the charged particle multiplicity is expected to be sensitive to $N_{\rm mpi}$, the correlation between  $flatenicity$ and the charged particle multiplicity is shown in Fig.~\ref{fig:5}. The behavior is qualitatively similar to that from the analysis as a function of $N_{\rm mpi}$.

\section{Conclusions}

In this work we have shown that PYTHIA~8.244 predicts the existence of hedgehog events in pp collisions. These type of events were observed at TEVATRON energies, but their properties were never studied.  Hedgehog events are produced when a moderate number of several semi-hard scatterings occur within the same pp collision. We observed a clear spreading of low-$p_{\rm T}$ particles in the whole $\eta-\varphi$ range. In this contribution we have also introduced a tool which can be used by experiments in order to tag this type of rare pp collisions. We have introduced {$flatenicity$}, we argue that this new event shape would allow to characterize the hedgehog events. This would requires an apparatus like ALICE 3 which is expected to have excellent particle identification capabilities at mid-pseudorapidity, as well as tracking over  a wide pseudorapidity range, and a low transverse momentum threshold.  These events, where one does not observe clear jet structures, could be an interesting subject for further investigation.

\section{Acknowledgement}
This work has been supported by CONACyT under the Grants CB No. A1-S-22917 and CF No. 2042. A. O. acknowledges the support received from UNAM under the program PASPA-DGAPA.

\section*{References}

\bibliographystyle{utphys}
\bibliography{biblio}

\end{document}